\documentclass[oneside,english]{elsart}
\usepackage[T1]{fontenc}
\usepackage[latin9]{inputenc}
\usepackage{amsmath}
\usepackage{amssymb}

\makeatletter
\DeclareMathOperator{\op}{op}
\DeclareMathOperator{\opt}{opt}
\DeclareMathOperator{\esssup}{esssup}
\DeclareMathOperator{\essinf}{essinf}
\DeclareMathOperator{\cls}{closure}
\DeclareMathOperator{\spa}{span}

\makeatother

\usepackage{babel}

\begin{document}
\begin{frontmatter}

\title{Zero-Forcing Precoding for Frequency Selective MIMO Channels with $H^\infty$ Criterion and Causality Constraint\thanksref{DFG}}

\thanks[DFG]{This work was supported by the German Research Foundation (DFG) under grants BO 1734/11-1 and BO 1734/11-2}

\author{Sander Wahls\corauthref{cor1}}
\ead{sander.wahls@mk.tu-berlin.de}
\corauth[cor1]{Corresponding author. Fon: +49 30 314 28559. Fax: +49 30 314 28320.}
\author{Holger Boche}
\ead{holger.boche@mk.tu-berlin.de}
\address{Technische Universität Berlin, Heinrich-Hertz-Lehrstuhl für Mobilkommunikation, Werner-von-Siemens-Bau (HFT6), Einsteinufer 25, 10587 Berlin, Germany}
\author{Volker Pohl}
\ead{pohl@ee.technion.ac.il}
\address{Technion - Israel Institute of Technology, Department of Electrical Engineering, Haifa 32000, Israel}

\begin{abstract}
We consider zero-forcing equalization of frequency selective MIMO channels by causal and linear time-invariant precoders in the presence of intersymbol interference. Our motivation is twofold. First, we are concerned with the  optimal performance of causal precoders from a worst case point of view. Therefore we construct an optimal causal precoder, whereas contrary to other works our construction is not limited to finite   or rational  impulse responses. Moreover we derive a novel numerical approach to computation of the optimal perfomance index achievable by causal precoders for given channels. This quantity is important in the numerical determination of optimal precoders.
\end{abstract}

\begin{keyword}
MIMO \sep Intersymbol Interference \sep Filterbank \sep Precoder \sep Equalizer \sep Causality \sep Bezout Identity \sep Matrix Corona Problem \sep Minimum Norm
\end{keyword}

\end{frontmatter}

\section{Introduction}

Many of todays state-of-the-art wireless systems adopt multiple-input
multiple-output (MIMO) transmission to increase spectral efficiency
together with multi-carrier methods to cope with intersymbol interference
(ISI). Multi-carrier methods simplify channel equalization because
they decompose frequency selective channels into multiple flat fading
channels (the so called carriers), which can be easily equalized.
While multi-carrier transmission offers many advantages including
effective channel equalization, it also exhibits some drawbacks regarding
the peak-to-average power ratio (PAPR). Often single-carrier transmission,
where the frequency selective channel is approached directly, is considered
as an alternative to multi-carrier transmission \cite{Falconer2002,Fischer2003,MLG1}.
While therefore single-carrier transmission is interesting on its
own, it has been further shown in \cite{SGB1,SY}, that in fact many
common multi-carrier, code-multiplex and space-time block-code systems
can be modeled as single-carrier systems by virtual enhancement of
the MIMO system. Various authors used this approach to derive new
equalization methods based on single-carrier equalization in order
to exploit joint equalization of spatial, time and code or frequency
domains \cite{SGB1,SY,GuLi1,GB1}. There, and generally for linear
time-invariant (LTI) equalization of single-carrier systems with zero-forcing
and causality constraint, one usually solves the so-called \emph{Bezout
Identity}\[
H(e^{i\theta})G(e^{i\theta})=I\qquad(0\le\theta<2\pi),\]
where the matrix-valued transfer function $H$ of a stable and causal
LTI system (the frequency selective MIMO channel) is given, and a
transfer function $G$ of a stable and causal LTI precoder, which
equalizes $H$, has to be computed \cite{WB}. Transmitters may use
such $G$ to pre-equalize the channel. Alternatively, receivers can
also solve the Bezout Identity for the transposed channel $H^{T}$
(i.e. $H^{T}G=I$) and equalize the channel $H$ with the transposed
solution $G^{T}$. The main difficulty in solving the Bezout Identity
is the causality of $G$, because the naive approach \[
G(e^{i\theta})=H(e^{i\theta})^{*}[H(e^{i\theta})H(e^{i\theta})^{*}]^{-1}\qquad(0\le\theta<2\pi)\]
of a \emph{pseudoinverse} generally results in a non-causal precoder
\cite{BP06cewp}. If the number of channel inputs equals the number
of channel outputs, the pseudoinverse is the unique solution to the
Bezout Identity. The situation changes if the number of inputs of
$H$ is larger than the number of outputs. Now precoders for $H$
no longer have to be unique. Usually this non-uniqueness then is exploited
to choose a causal $G$ that is optimal in some sense. The two common
optimality conditions are \emph{minimality of the equalizers energy}
and \emph{minimality of the equalizers peak value}, respectively.
The minimal energy condition corresponds to the classical approach
of signal-to-noise ratio (SNR) maximization \cite{SY,GB1,Li2005}.
However, this approach is only feasible if the statistical properties
of the noise are known. For unknown noise statistics, it cannot be
applied. Picking up an idea from robust control (see e.g. \cite{ZCG96}),
where one is concerned with unpredictable errors that arise e.g. due
to uncertain modeling, instead minimization of the equalizers peak
value has been proposed \cite{GuLi1}. As we shall see later, this
minimizes the worst case error instead of the average error, which
cannot be determined due to the unknown noise statistics.

In this paper we are interested in the optimal performance that causal
precoders and equalizers can archive regarding the worst case error.
Therefore we show how a solution to the Bezout Identity with minimal
peak value can be constructed. We discuss why this gives the best
upper bounds on various perturbations in the system. Contrary to other
ways to solve the Bezout Identity, our construction holds for the
most general case of systems with infinite impulse responses (which
are not required to be rational) and even infinite input and output
vectors, i.e. we allow systems to have infinite temporal as well as
infinite spatial dimension. We further give a new result on the numerical
computation of the minimum peak value achievable by causal solutions
to the Bezout Identity if the numbers of inputs and outputs are finite.
This is important because for all methods known to the authors that
solve the Bezout Identity with minimal peak value in a numerically
exploitable way, the minimal peak value has to be known in advance
\cite{GuLi1,Tr4}. Therefore efficient computation of the minimal
peak value is important for numerical solution of the Bezout Identity.
We point out that the optimization approach in \cite{VB06efmi} requires
no prior knowledge of the minimal peak value. However, it only computes
finite impulse response solutions to the Bezout Identity, which are
generally suboptimal. 

We proceed as follows. In Section 2 we give our problem statement
after we have introduced some notation and necessary basic mathematical
concepts. We further discuss the practical interpretation of our problem
statement. In Section 3 we derive our results on the numerical computation
of the minimal peak value achievable by causal solutions to the Bezout
Identity. Then a optimal causal precoder is constructed in Section
4. We finally draw conclusions in Section 5.

\section{Preliminaries}

\subsection{Notation}

We denote the complex numbers by $\mathbb{C}$, the complex matrices
with $m$ rows and $n$ columns by $\mathbb{C}^{m\times n}$ and the
complex column vectors by $\mathbb{C}^{m}:=\mathbb{C}^{m\times1}$.
The complex unit disc is given as $\mathbb{D}:=\{z\in\mathbb{C}:|z|<1\}$,
its border is the unit circle $\mathbb{T}:=\{z\in\mathbb{C}:|z|=1\}$.
Complex conjugation is denoted by $\bar{(\cdot)}$, taking adjoints
in a Hilbert space by $(\cdot)^{*}$. Furthermore $\mathcal{H}$,$\mathcal{E}$
and $\mathcal{E}_{*}$ denote separable Hilbert spaces with scalar
products $\langle\cdot,\cdot\rangle_{\mathcal{H}}$, $\langle\cdot,\cdot\rangle_{\mathcal{E}}$
and $\langle\cdot,\cdot\rangle_{\mathcal{E}_{*}}$, respectively.
By $\mathcal{H}\oplus\mathcal{E}$ we mean the direct Hilbert sum,
i.e. the space $\mathcal{H}\times\mathcal{E}$ equipped with scalar
product $\langle h\oplus e,g\oplus f\rangle_{\mathcal{H}\oplus\mathcal{E}}:=\langle h,g\rangle_{\mathcal{H}}+\langle e,f\rangle_{\mathcal{E}}$.
The space of bounded linear operators between $\mathcal{E}$ and $\mathcal{E}_{*}$
is denoted by $\mathcal{L}(\mathcal{E},\mathcal{E}_{*})$. It is equipped
with the operator norm $\Vert T\Vert_{\op}:=\sup_{e\in\mathcal{E},\Vert e\Vert_{\mathcal{E}}=1}\Vert Te\Vert_{\mathcal{E}_{*}}$.
On any space the identity operator is written as $I$. For matrices
$A\in\mathbb{C}^{m\times n}$ the smallest and largest singular value
will be denoted by $\sigma_{\min}(A)$ and $\sigma_{\max}(A)$, respectively.
The closure of a set $M$ is denoted by $\cls M$, the space spanned
by all linear combinations of its elements by $\spa M$.

As usual, $L_{\mathbb{T}}^{p}(X)$ denotes the space of (equivalence
classes of) $p$-integrable functions on $\mathbb{T}$ with values
in a Banach space $X$. The norm in $L_{\mathbb{T}}^{p}(X)$ is $\Vert f\Vert_{p}^{p}:=\int_{\theta=0}^{2\pi}\Vert f(e^{i\theta})\Vert_{X}^{p}\frac{d\theta}{2\pi}$
for $1\le p<\infty$ and $\Vert f\Vert_{\infty}:=\esssup_{\zeta\in\mathbb{T}}\Vert f(\zeta)\Vert_{X}=\inf\left\{ m>0:\mu(\{\zeta\in\mathbb{T}:\Vert f(\zeta)\Vert_{X}>m\})=0\right\} $
for $p=\infty$, where $\mu$ denotes the Lebesgue measure. We refer
to \cite[Section 3.11]{Nik1} and the references therein for details
on integration of vector- and operator-valued functions. If $p=2$,
$L_{\mathbb{T}}^{2}(\mathcal{E})$ equipped with the scalar product
$\langle f,g\rangle_{2}:=\int_{\theta=0}^{2\pi}\langle f(e^{i\theta}),g(e^{i\theta})\rangle_{\mathcal{E}}\frac{d\theta}{2\pi}$
is a Hilbert space. For $F\in L_{\mathbb{T}}^{\infty}(\mathcal{L}(\mathcal{E},\mathcal{E}_{*}))$
we denote the point-wise adjoint by $F^{*}$, i.e. $F^{*}(\zeta)=(F(\zeta))^{*}$
almost everywhere on the unit circle.

\subsection{Basic Results and Concepts}

\subsubsection{Hardy Spaces and Toeplitz Operators}

We introduce the usual \emph{Hardy spaces on the disc} by\begin{eqnarray*}
H_{\mathbb{D}}^{2}(\mathcal{E}) & := & \left\{ u:\mathbb{D}\to\mathcal{E}:u\text{ analytic},\Vert u\Vert_{2}^{2}:=\sup_{0<r<1}\int_{\theta=0}^{2\pi}\Vert u(re^{i\theta})\Vert_{\mathcal{E}}^{2}\frac{d\theta}{2\pi}<\infty\right\} ,\\
H_{\mathbb{D}}^{\infty}(\mathcal{E},\mathcal{E}_{*}) & := & \left\{ F:\mathbb{D}\to\mathcal{L}(\mathcal{E},\mathcal{E}_{*}):F\text{ analytic},\Vert F\Vert_{\infty}:=\sup_{z\in\mathbb{D}}\Vert F(z)\Vert_{\op}<\infty\right\} .\end{eqnarray*}
The Hardy spaces play an important role in systems theory, since they
are the set of transfer functions of causal finite energy signals
and causal and energy-stable transfer functions for LTI systems, respectively
\cite{BP05gsot}. Definition is also possible on the upper half plane
instead of the unit disc. On both domains, the Hardy functions are
completely determined by their values on the borders of the domain.
Therefore, each Hardy function on the unit disc has a corresponding
function on the circle. The space of those corresponding functions
can be given as follows.

For functions $f\in L_{\mathbb{T}}^{1}(\mathcal{E})$ or $f\in L_{\mathbb{T}}^{1}(\mathcal{L}(\mathcal{E},\mathcal{E}_{*}))$
the $k$-th \emph{Fourier coefficient} is \[
\hat{f}_{k}:=\int_{\theta=0}^{2\pi}f(e^{i\theta})e^{-ik\theta}\frac{d\theta}{2\pi}\qquad(k\in\mathbb{Z}).\]
Therewith, the \emph{Hardy spaces on the circle} are given by\begin{eqnarray*}
H_{\mathbb{T}}^{2}(\mathcal{E}) & := & \left\{ u\in L_{\mathbb{T}}^{2}(\mathcal{E}):\hat{u}_{k}=0\text{ for }k<0\right\} ,\\
H_{\mathbb{T}}^{\infty}(\mathcal{E},\mathcal{E}_{*}) & := & \left\{ F\in L_{\mathbb{T}}^{\infty}(\mathcal{L}(\mathcal{E},\mathcal{E}_{*})):\hat{F}_{k}=0\text{ for }k<0\right\} .\end{eqnarray*}

It is important to know that the two notions of Hardy spaces on disc
and circle are equivalent, since by Fatou's Theorem the radial limit
$(bu)(e^{i\theta}):=\lim_{r\nearrow1}u(re^{i\theta})$ exists almost
everywhere and  the mapping $b$ is an isometric isometry between
the Hardy spaces on disc and circle (see \cite[Th. 3.11.7, 3.11.10]{Nik1}).
Therefore we will only explicitly distinguish between those spaces
if necessary, and simply write $H^{2}(\mathcal{E})$ and $H^{\infty}(\mathcal{E},\mathcal{E}_{*})$
otherwise. 

An important property of $L_{\mathbb{T}}^{2}(\mathcal{E})$ is \emph{Parseval's
Relation} (\cite[p. 184]{SzNF}), by which\[
\Vert u\Vert_{2}^{2}=\sum_{k=-\infty}^{\infty}\Vert\hat{u}_{k}\Vert_{\mathcal{E}}^{2}\text{ for all }u\in L_{\mathbb{T}}^{2}(\mathcal{E}).\]

We will now introduce \emph{Toeplitz operators}, which are the standard
example for operators on Hardy spaces and which will also play an
important role in what follows. The orthogonal projection $(P_{+}u)(\zeta):=\sum_{k=0}^{\infty}\hat{u}_{k}\zeta^{k}$
from $L_{\mathbb{T}}^{2}(\mathcal{E})$ into $H_{\mathbb{T}}^{2}(\mathcal{E})$
is called the \emph{Riesz Projection}. The projection from $H_{\mathbb{T}}^{2}(\mathcal{E})$
into the space of degree $N$ polynomials is $(P_{N}u)(\zeta):=\sum_{k=0}^{N}\hat{u}_{k}\zeta^{k}$.
Now for $F\in L_{\mathbb{T}}^{\infty}(\mathcal{E},\mathcal{E}_{*})$
the \emph{Toeplitz operator with symbol $F$} is the operator which
maps $H_{\mathbb{T}}^{2}(\mathcal{E})$ into $H_{\mathbb{T}}^{2}(\mathcal{E}_{*})$
via $T_{F}u:=P_{+}(Fu)$. 

The next result allows us to get an exact estimate of the minimum
norm achievable by solutions of the Bezout Identity.

\begin{thm}
[{\cite[Th. 9.2.1]{Nik1}}]\label{thm:toeplitz-corona}Let $F\in H^{\infty}(\mathcal{E},\mathcal{E}_{*})$
and $\delta>0$. Then some $G\in H^{\infty}(\mathcal{E}_{*},\mathcal{E})$
with $\Vert G\Vert_{\infty}\le\delta^{-1}$ and $F(z)G(z)=I$ for
all $z\in\mathbb{D}$ exists if and only if\[
\Vert T_{F^{*}}u\Vert_{2}\ge\delta\Vert u\Vert_{2}\text{ for all }u\in H^{2}(\mathcal{E}_{*}).\]

\end{thm}

\subsubsection{Schur Class}

Functions in the unit ball of $H^{\infty}(\mathcal{E},\mathcal{E}_{*})$,
the so-called \emph{Schur class} \[
S(\mathcal{E},\mathcal{E}_{*}):=\left\{ F\in H^{\infty}(\mathcal{E},\mathcal{E}_{*}):\Vert F\Vert_{\infty}\le1\right\} ,\]
have some special properties, which will turn out to be useful in
the construction of a minimum norm right inverse. Every Schur function
can be factorized as follows.

\begin{thm}
[{\cite[Th. 2.1]{BaTr1}}]\label{thm:agler}Let $F:\mathbb{D}\to\mathcal{L}(\mathcal{E},\mathcal{E}_{*})$.
Then $F\in S(\mathcal{E},\mathcal{E}_{*})$ if and only if there exists
a holomorphic function $W:\mathbb{D}\to\mathcal{L}(\mathcal{H},\mathcal{E}_{*})$
such that\[
I-F(z)F(w)^{*}=(1-z\bar{w})W(z)W(w)^{*}\qquad(z,w\in\mathbb{D}).\]

\end{thm}
Note that $W$ can be given explicitly, see \cite[Sec. 3.3]{BaTr1}.
We finish with the observation that also certain block operators define
Schur functions. 

\begin{lem}
[{\cite[Lem. 2]{Tr4}}]\label{lem:sz-nagy-foias}Let $T\in\mathcal{L}(\mathcal{H}\oplus\mathcal{E},\mathcal{H}\oplus\mathcal{E}_{*})$
with $\Vert T\Vert_{\op}\le1$. Then $T$ has a unique block matrix
representation\[
T=\left[\begin{array}{cc}
A & B\\
C & D\end{array}\right]:\left[\begin{array}{c}
\mathcal{H}\\
\mathcal{E}\end{array}\right]\to\left[\begin{array}{c}
\mathcal{H}\\
\mathcal{E}_{*}\end{array}\right]\]
and the function\[
F:\mathbb{D}\to\mathcal{L}(\mathcal{E},\mathcal{E}_{*}),\qquad F(z):=D+Cz(I-zA)^{-1}B\]
is Schur, i.e. $F\in S(\mathcal{E},\mathcal{E}_{*})$. 
\end{lem}
Functions defined as $F$ in the Lemma above are known in operator
theory as characteristic functions, while unitary operators like $T$
are known as unitary colligations. Those concepts resemble much the
concept of a transfer function and a state-space realization in control
theory. We refer to \cite{BaTr1,Br1} for details.

\subsection{Problem Formulation}

Before we give an exact problem formulation we introduce and discuss
the target objective\[
\gamma_{\opt}(H):=\inf\left(\left\{ \Vert G\Vert_{\infty}:G\in H^{\infty}(\mathcal{E}_{*},\mathcal{E}),H(z)G(z)=I\text{ for all }z\in\mathbb{D}\right\} \cup\{\infty\}\right),\]
which is, as we will see, a tight lower bound on the worst-case transmit
energy enhancement of causal precoders for the channel $H$, and a
measure for the achievable robustness against imperfectly known channel
transfer functions. Note that in particular $\gamma_{\opt}(H)=\infty$
if and only if $H$ has no right inverse in $H^{\infty}$. We always
assume $H\in H^{\infty}(\mathcal{E},\mathcal{E}_{*})$ unless we explicitly
mention otherwise.

It was shown in \cite{Tol1} that if $\dim\mathcal{E}_{*}<\infty$,
existence of a right inverse in $H^{\infty}$ is further equivalent
to \[
H(z)H(z)^{*}\ge\delta^{2}I\text{ for some }\delta>0\text{ and all }z\in\mathbb{D}.\]
It is somewhat surprising that although by the result from \cite{Tol1}
$\gamma_{\opt}<\infty$ if and only if \[
\delta_{c}:=\sup\left\{ \delta\ge0\mid H(z)H(z)^{*}\ge\delta^{2}I\text{ for all }z\in\mathbb{D}\right\} >0,\]
$\delta_{c}$ has no direct connection to $\gamma_{\opt}$, i.e. $\gamma_{\opt}$
cannot be computed from $\delta_{c}$ \cite{BP06cewp}. However, as
we will see, it is important to know $\gamma_{\opt}$ in advance of
the construction of an optimal precoder. Therefore we derive a new
method for numerical computation of $\gamma_{\opt}$ and then solve
the following problem. 

\begin{prob}
\label{pro:min-corona}Let $\gamma_{\opt}(H)<\infty$. How can $G\in H^{\infty}(\mathcal{E}_{*},\mathcal{E})$
with $H(z)G(z)=I$ for all $z\in\mathbb{D}$ and $\Vert G\Vert_{\infty}=\gamma_{\opt}(H)$
be constructed?
\end{prob}
We close this section with a short discussion in which sense minimization
of the infinity norm in Problem \ref{pro:min-corona} gives optimal
filters. The input-output relation of a frequency selective MIMO channel
is given by\[
y(\zeta)=H(\zeta)x(\zeta)+n(\zeta)\qquad(\zeta\in\mathbb{T}),\]
where $H$ denotes the channel, $x$ the transmitted signals and $y$
and $n$ the received signals and additive noise, respectively. If
a precoder $G$ for $H$ is used to pre-distort the transmitted signals,
this input-output relation changes to\[
y(\zeta)=H(\zeta)G(\zeta)x(\zeta)+n(\zeta)=x(\zeta)+n(\zeta)\qquad(\zeta\in\mathbb{T}).\]
 There are two advantages in minimizing the infinity norm of the precoder.

The first advantage is minimization of the transmit signals energy.
The energy necessary to transmit a signal $x$ using the precoder
$G$ is given by $\Vert Gx\Vert_{2}^{2}$. Without loss of generality,
let us assume that $\Vert x\Vert_{2}^{2}=1$. Then, it can be shown
that the transmit energy necessary in the worst case is exactly $\Vert G\Vert_{\infty}^{2}$,
i.e. \[
\sup_{x\in H^{2}(\mathcal{E}_{*}),\Vert x\Vert_{2}=1}\Vert Gx\Vert_{2}^{2}=\Vert G\Vert_{\infty}^{2}.\]
Thus, minimizing $\Vert G\Vert_{\infty}$ guarantees the lowest amount
of necessary transmit energy. If equalizers instead of precoders are
considered, i.e.\[
y(\zeta)=G(\zeta)[H(\zeta)x(\zeta)+n(\zeta)]=x(\zeta)+G(\zeta)n(\zeta)\qquad(\zeta\in\mathbb{T}),\]
this is equivalent to minimal worst case noise enhancement.

The second advantage of minimization of the infinity norm is robustness.
Assume an imperfectly known channel transfer function $H_{\Delta}=H+\Delta$
with right inverse $G_{\Delta}$, where $H$ is the correct channel
and $\Delta$ is a perturbation. Using the same argument as before,
we see that the energy of the worst error that can result from the
perturbation equals\[
\sup_{x\in H^{2}(\mathcal{E}_{*}),\Vert x\Vert_{2}=1}\Vert x-HG_{\Delta}x\Vert_{2}^{2}=\sup_{x\in H^{2}(\mathcal{E}_{*}),\Vert x\Vert_{2}=1}\Vert\Delta G_{\Delta}x\Vert_{2}^{2}=\Vert\Delta G_{\Delta}\Vert_{\infty}^{2}.\]
Since it holds $\Vert\Delta G_{\Delta}\Vert_{\infty}^{2}\le\Vert\Delta\Vert_{\infty}^{2}\Vert G_{\Delta}\Vert_{\infty}^{2}$,
and this inequality can become sharp e.g. for $\Delta=\delta I$,
we see that minimizing $\Vert G_{\Delta}\Vert_{\infty}$ also minimizes
the worst case error that results from an imperfectly known channel
transfer function. This argument applies to equalizers in the same
way it applies to precoders.

\section{Computation of the Optimal Norm}

This section deals with the computation of the optimal norm $\gamma_{\opt}$
achievable by solutions to the Bezout Identity. Since many algorithms
which directly solve Problem \ref{pro:min-corona} only compute suboptimal
solutions, i.e. given $\gamma>\gamma_{\opt}$ they compute a right
inverse $G_{\gamma}$ with norm $\Vert G_{\gamma}\Vert_{\infty}<\gamma$,
it is important to know the optimal value for $\gamma$ in advance
\cite{GuLi1,Tr4}. We point out that computation of $\gamma_{\opt}$
also arises in other contexts, see e.g. Remark 1 in \cite{GS1} (with
the next corollary in mind).

We start with an exact (but incomputable) formula for $\gamma_{\opt}$.
The next two corollaries are direct consequences of Theorem \ref{thm:toeplitz-corona}.

\begin{cor}
\label{cor:gamma-opt-is-one-div-rho}For $\rho(H):=\inf_{u\in H^{2}(\mathcal{E}_{*}),\Vert u\Vert_{2}=1}\Vert T_{H^{*}}u\Vert_{2}$,
it holds $\gamma_{\opt}(H)=\rho(H)^{-1}$.
\begin{cor}
\label{cor:min-norm-right-inverse-exists}If $\gamma_{\opt}(H)<\infty$,
a right inverse $G\in H^{\infty}(\mathcal{E}_{*},\mathcal{E})$ with
$\Vert G\Vert_{\infty}=\gamma_{\opt}(H)$ exists.
\end{cor}
\end{cor}
The interesting thing about Corollary \ref{cor:gamma-opt-is-one-div-rho}
is that it shows us why the optimal causal equalizer cannot perform
better than the optimal non-causal one. Note that the optimal norm
for non-causal equalizers is given by \[
\left(\inf_{u\in H^{2}(\mathcal{E}_{*}),\Vert u\Vert_{2}=1}\Vert H^{*}u\Vert_{2}\right)^{-1}\]
 (see \cite{BP06cewp}), which is the same formula as Corollary \ref{cor:gamma-opt-is-one-div-rho},
except for the additional Riesz projection $P_{+}$:\[
\gamma_{\opt}(H)=\left(\inf_{u\in H^{2}(\mathcal{E}_{*}),\Vert u\Vert_{2}=1}\Vert P_{+}(H^{*}u)\Vert_{2}\right)^{-1}.\]
It is now clear that causal equalizers perform worse because the signal
energy of $u$ which is mapped into the non-causal part of $H^{*}u$
is cut off. How much energy is shifted into the non-causal part thereby
depends on the Fourier coefficients of $H^{*}$, which are related
to $H$ by $\widehat{H^{*}}_{k}=\hat{H}_{-k}^{*}$ for $k\in\mathbb{Z}$.

We now derive a computable approximation of $\gamma_{\opt}$. The
main idea will be to approximate the relation $\gamma_{\opt}=\rho^{-1}$
from Corollary \ref{cor:gamma-opt-is-one-div-rho}. In order to compute
$\gamma_{\opt}$, we try to approximate $\rho$ with\[
\rho_{N}(H):=\inf_{u\in P_{N}H^{2}(\mathcal{E}_{*}),\Vert u\Vert_{2}=1}\Vert P_{N}T_{H^{*}}u\Vert_{2},\]
i.e. we restrict domain and image of $T_{H^{*}}$ to polynomials of
degree $N$ and take the infimum for this restriction. Because $P_{N}T_{H^{*}}P_{N}$
is linear and finite dimensional, it can be represented by a matrix.

The main result of this section is the following.

\begin{thm}
\label{thm:rho-N-converges}The sequence $\{\rho_{N}(H)\}_{N}$ is
monotonically decreasing  and converges with limit \[
\lim_{N\to\infty}\rho_{N}(H)=\rho(H)=\gamma_{\opt}(H)^{-1}.\]
If $H\in H^{\infty}(\mathbb{C}^{m\times n})$ with $m\le n$,%
\footnote{Note that trivially $\gamma_{\opt}(H)=\infty$ for $m>n$.%
} and \[
\Gamma_{H,N}:=\left[\begin{array}{cccc}
\hat{H}_{0}^{*} & \hat{H}_{1}^{*} & \dots & \hat{H}_{N}^{*}\\
0 & \hat{H}_{0}^{*} & \dots & \hat{H}_{N-1}^{*}\\
\vdots & \ddots & \ddots & \vdots\\
0 & \dots & 0 & \hat{H}_{0}^{*}\end{array}\right]\in\mathbb{C}^{n(N+1)\times m(N+1)},\]
$\rho_{N}$ can be computed as $\rho_{N}(H)=\sigma_{\min}(\Gamma_{H,N})$.
\end{thm}
\begin{pf}
We only sketch the proof here, the full proof is given in the appendix.
It consists of three main steps. The first step is to show that the
sequence $\{\rho_{N}(H)\}_{N}$ is monotonically decreasing and lower
bounded by $\rho(H)$. The main idea is that the relation \[
\rho_{N}(H)=\inf_{u\in P_{N}H^{2}(\mathcal{E}_{*}),\Vert u\Vert_{2}=1}\Vert P_{N}T_{H^{*}}u\Vert_{2}=\inf_{u\in P_{N}H^{2}(\mathcal{E}_{*}),\Vert u\Vert_{2}=1}\Vert T_{H^{*}}u\Vert_{2}\]
holds for every $N\in\mathbb{N}$ and thus the infimum is always taken
over the same target objective, but over a space which increases with
$N$. This is done in the appendix in Proposition \ref{pro:rho-k-mon-dec}.
In a second step it is shown that the lower bound $\rho(H)$ for $\{\rho_{N}(H)\}_{N}$
is sharp. Therefore for arbitrary $\epsilon>0$ a sequence $\{u_{N}\}_{N}$
such that\[
u_{N}\in P_{N}H^{2}(\mathcal{E}_{*}),\Vert u_{N}\Vert_{2}=1\text{ and }\lim_{N\to\infty}\Vert P_{N}T_{H^{*}}u_{N}\Vert_{2}\le\rho(H)+\epsilon\]
is constructed in the appendix in Proposition \ref{pro:rho-k-le-rho-plus-eps}.
Thus $\rho_{N}(H)$ converges to $\rho(H)$, which is equal to $\gamma_{\opt}(H)^{-1}$
by Corollary \ref{cor:gamma-opt-is-one-div-rho}. Finally Proposition
\ref{pro:rho-N-is-sigma-min} in the appendix gives the formula for
computation of $\rho_{N}(H)$ via singular value decomposition if
$H$ is matrix-valued.$\hfill\blacksquare$
\end{pf}
Since the arguments used to prove Theorem \ref{thm:rho-N-converges}
hold analogously if we approximate \[
\sup_{u\in H^{2}(\mathbb{C}^{m}),\Vert u\Vert_{2}=1}\Vert T_{H^{*}}u\Vert_{2}=\Vert T_{H^{*}}\Vert_{\op}=\Vert T_{H}^{*}\Vert_{\op}=\Vert T_{H}\Vert_{\op}=\Vert H\Vert_{\infty}\]
instead of $\rho_{N}(H)=\inf_{u\in H^{2}(\mathbb{C}^{m}),\Vert u\Vert_{2}=1}\Vert T_{H^{*}}u\Vert_{2}$,
we also see that for $H\in H^{\infty}(\mathbb{C}^{m\times n})$ the
sequence $\{\sigma_{\max}(\Gamma_{H,N})\}_{N}$ is monotonically increasing
and converges with limit \[
\lim_{N\to\infty}\sigma_{\max}(\Gamma_{H,N})=\Vert H\Vert_{\infty}.\]
We note that the well-known fact that the limit $\Vert H\Vert_{\infty}$
of $\sigma_{\max}(\Gamma_{H,N})$ can be found by performing a grid
search over all frequencies, i.e. \[
\lim_{N\to\infty}\sigma_{\max}(\Gamma_{H,N})=\Vert H\Vert_{\infty}=\esssup_{\zeta\in\mathbb{T}}\sigma_{\max}(H(\zeta)),\]
does not carry over to computation of $\gamma_{\opt}(H)$. Here, in
general we have \[
\lim_{N\to\infty}\sigma_{\min}(\Gamma_{H,N})=\gamma_{\opt}(H)^{-1}\ne\essinf_{\zeta\in\mathbb{T}}\sigma_{\min}(H(\zeta)).\]
This dichotomy results from the fact that while indeed \[
\sup_{u\in H^{2}(\mathcal{E}_{*}),\Vert u\Vert_{2}=1}\Vert H^{*}u\Vert_{2}=\sup_{u\in H^{2}(\mathcal{E}_{*}),\Vert u\Vert_{2}=1}\Vert P_{+}(H^{*}u)\Vert_{2},\]
 in general we have \[
\inf_{u\in H^{2}(\mathcal{E}_{*}),\Vert u\Vert_{2}=1}\Vert H^{*}u\Vert_{2}\ne\inf_{u\in H^{2}(\mathcal{E}_{*}),\Vert u\Vert_{2}=1}\Vert P_{+}(H^{*}u)\Vert_{2}.\]
This can be easily seen in the next example.

\begin{exmp}
Set $H(\zeta)=\zeta$ for $\zeta\in\mathbb{T}$. Then by Parseval's
Relation \[
\inf_{u\in H^{2}(\mathbb{C}),\Vert u\Vert_{2}=1}\Vert H^{*}u\Vert_{2}=\inf_{u\in H^{2}(\mathbb{C}),\Vert u\Vert_{2}=1}\Vert u\Vert_{2}=1,\]
however for $u(z)=1$ we have $(H^{*}u)(\zeta)=\bar{\zeta}$ and therefore
\[
\Vert P_{+}(H^{*}u)\Vert_{2}=\Vert0\Vert_{2}=0.\]

\end{exmp}
It is also important to note that Theorem \ref{thm:rho-N-converges}
does not generalize to the case $H\in L_{\mathbb{T}}^{\infty}$. We
give an example where $\rho(H)=1$, a inverse in $H^{\infty}$ exists,
but the smallest singular values of the finite sections do not converge
to $\rho(H)$.

\begin{exmp}
Set $H(\zeta):=\bar{\zeta}$ for $\zeta\in\mathbb{T}$. Then by Parseval's
Relation\[
\rho(H)=\inf_{u\in H^{2}(\mathbb{C}),\Vert u\Vert_{2}=1}\Vert T_{H^{*}}u\Vert_{2}=\inf_{u\in H^{2}(\mathbb{C}),\Vert u\Vert_{2}=1}\Vert T_{\zeta}u\Vert_{2}=\inf_{u\in H^{2}(\mathbb{C}),\Vert u\Vert_{2}=1}\Vert u\Vert_{2}=1.\]
Further, $H$ has a inverse in $H^{\infty}$, i.e. $G(\zeta)=\zeta$.
However,

\[
\sigma_{\min}\left(\left[\begin{array}{cccc}
\hat{H}_{0}^{*} & \hat{H}_{1}^{*} & \dots & \hat{H}_{N}^{*}\\
\hat{H}_{-1}^{*} & \hat{H}_{0}^{*} & \ddots & \vdots\\
\vdots & \ddots & \ddots & \hat{H}_{1}^{*}\\
\hat{H}_{-N}^{*} & \dots & \hat{H}_{-1}^{*} & \hat{H}_{0}^{*}\end{array}\right]\right)=\sigma_{\min}\left(\left[\begin{array}{cccc}
0 & \dots & \dots & 0\\
1 & \ddots &  & \vdots\\
 & \ddots & \ddots & \vdots\\
 &  & 1 & 0\end{array}\right]\right)=0\]
for all $N\in\mathbb{N}$.
\end{exmp}

\section{Construction of the Optimal Causal Precoder}

In this section we construct a minimum norm solution to the Bezout
Identity, i.e. we solve Problem \ref{pro:min-corona}. The major idea
of the proof is the following. We first show how to construct right
inverses with norm at most one. Then given any $H\in H^{\infty}(\mathcal{E},\mathcal{E}_{*})$,
we apply this technique to the scaled function $\gamma_{\opt}H$.
Appropriate rescaling of the obtained inverse will result in a minimum
norm right inverse.

\subsection{Schur Right Inverse}

The first step is construction of a Schur right inverse. Therefore
we factorize the function to be inverted similar to Theorem \ref{thm:agler}
and use this factorization to construct a contraction of the form
of $T$ in Lemma \ref{lem:sz-nagy-foias}. The characteristic function
of this contraction then is the wanted right inverse. This is a variant
of the technique known as {}``lurking isometry method'', which has
been introduced by Ball and Trent \cite[Th. 5.2]{BaTr1} and independently
Agler and McCarthy \cite{AgMc} to solve the Bezout Identity.

We start with the factorization.

\begin{lem}
\label{lem:decomposition-of-F}Let $H$ have a right inverse $G\in S(\mathcal{E}_{*},\mathcal{E})$.
Then there exits a holomorphic function $W:\mathbb{D}\to\mathcal{L}(\mathcal{H},\mathcal{E}_{*})$
such that\begin{equation}
H(z)H(w)^{*}-I=(1-z\bar{w})W(z)W(w)^{*}\qquad(z,w\in\mathbb{D}).\label{eq:lem:decomposition-of-F}\end{equation}

\end{lem}
\begin{pf}
By Theorem \ref{thm:agler} there exists a holomorphic function $\tilde{W}:\mathbb{D}\to\mathcal{L}(\tilde{\mathcal{H}},\mathcal{E})$
such that $I-G(z)G(w)^{*}=(1-z\bar{w})\tilde{W}(z)\tilde{W}(w)^{*}$.
Thus\[
H(z)H(w)^{*}-H(z)G(z)G(w)^{*}H(w)^{*}=(1-z\bar{w})H(z)\tilde{W}(z)\tilde{W}(w)^{*}H(w)^{*}.\]
Since $HG=I$ we obtain with $W(z):=H(z)\tilde{W}(z)$ that\[
H(z)H(w)^{*}-I=(1-z\bar{w})W(z)W(w)^{*}.\]
$\hfill\blacksquare$
\end{pf}
We can now introduce the appropriate block operator.

\begin{defn}
\label{def:V0}Let $H$ have a decomposition like \eqref{eq:lem:decomposition-of-F}
in Lemma \ref{lem:decomposition-of-F}. We define the sets\begin{eqnarray*}
D_{0} & := & \cls\left(\spa\left\{ \left[\begin{array}{c}
\bar{w}W(w)^{*}\\
H(w)^{*}\end{array}\right]e_{*}:w\in\mathbb{D},e_{*}\in\mathcal{E}_{*}\right\} \right)\subset\mathcal{H}\oplus\mathcal{E},\\
R_{0} & := & \cls\left(\spa\left\{ \left[\begin{array}{c}
W(w)^{*}\\
I\end{array}\right]e_{*}:w\in\mathbb{D},e_{*}\in\mathcal{E}_{*}\right\} \right)\subset\mathcal{H}\oplus\mathcal{E}_{*},\end{eqnarray*}
and a function $V_{0}:D_{0}\to R_{0}$ by\[
\sum_{k=0}^{\infty}c_{k}\left[\begin{array}{c}
\bar{w}W(w)^{*}\\
H(w)^{*}\end{array}\right]e_{*k}\mapsto\sum_{k=0}^{\infty}c_{k}\left[\begin{array}{c}
W(w)^{*}\\
I\end{array}\right]e_{*k}.\]

\end{defn}
Note that it can be easily shown with \eqref{eq:lem:decomposition-of-F}
that $V_{0}$ is a isometry, i.e. \[
\left\langle V_{0}\left[\begin{array}{c}
h\\
e\end{array}\right],V_{0}\left[\begin{array}{c}
h\\
e\end{array}\right]\right\rangle _{\mathcal{H}\oplus\mathcal{E}_{*}}=\left\langle \left[\begin{array}{c}
h\\
e\end{array}\right],\left[\begin{array}{c}
h\\
e\end{array}\right]\right\rangle _{\mathcal{H}\oplus\mathcal{E}}\text{ for all }\left[\begin{array}{c}
h\\
e\end{array}\right]\in\mathcal{H}\oplus\mathcal{E}.\]
Later we will use this fact when we apply Lemma \ref{lem:sz-nagy-foias}
to an extension of $V_{0}$. The wanted right inverse can now be given
explicitly.

\begin{thm}
\label{thm:schur-inverse}Let $H$ have a decomposition like \eqref{eq:lem:decomposition-of-F}
in Lemma \ref{lem:decomposition-of-F} and construct $V_{0}$ as in
Definition \ref{def:V0}. Denote by\[
V_{00}=\left[\begin{array}{cc}
A & B\\
C & D\end{array}\right]:\left[\begin{array}{c}
\mathcal{H}\\
\mathcal{E}\end{array}\right]\to\left[\begin{array}{c}
\mathcal{H}\\
\mathcal{E}_{*}\end{array}\right]\]
the continuation of $V_{0}$ with zero, i.e. \[
V_{00}d=\left\{ \begin{array}{cl}
V_{0}d & ,d\in D_{0}\\
0 & ,d\notin D_{0}\end{array}\right..\]
Then the function \[
G(z):=D^{*}+B^{*}(I-zA^{*})^{-1}zC^{*}\qquad(z\in\mathbb{D})\]
is a Schur right inverse of $H$, i.e. $G\in S(\mathcal{E}_{*},\mathcal{E})$
and $HG=I$.
\end{thm}
\begin{pf}
Let $w\in\mathbb{D}$. By construction of $V_{00}$ it holds\[
\left[\begin{array}{cc}
A & B\\
C & D\end{array}\right]\left[\begin{array}{c}
\bar{w}W(w)^{*}\\
H(w)^{*}\end{array}\right]e_{*}=\left[\begin{array}{c}
W(w)^{*}\\
I\end{array}\right]e_{*},\]
for all $e_{*}\in\mathcal{E}_{*}$, which is equivalent to\begin{equation}
A\bar{w}W(w)^{*}+BH(w)^{*}=W(w)^{*}\label{eq:thm:schur-inverse-1}\end{equation}
and\begin{equation}
C\bar{w}W(w)^{*}+DH(w)^{*}=I.\label{eq:thm:schur-inverse-2}\end{equation}
Since $\Vert V_{00}\Vert_{\op}\le\Vert V_{0}\Vert_{\op}=1$ because
$V_{0}$ is an isometry, we have $\Vert A\Vert_{\op}\le1$ and thus
$\Vert A\bar{w}\Vert_{\op}<1$. Thus $I-A\bar{w}$ is invertible,
and \eqref{eq:thm:schur-inverse-1} yields\[
W(w)^{*}=(I-A\bar{w})^{-1}BH(w)^{*}.\]
Plugging this representation of $W(w)^{*}$ into \eqref{eq:thm:schur-inverse-2}
results in\[
C\bar{w}(I-A\bar{w})^{-1}BH(w)^{*}+DH(w)^{*}=I.\]
Taking adjoints and replacing $w$ by $z$ shows that\[
H(z)\left[D^{*}+B^{*}(I-zA^{*})^{-1}zC^{*}\right]=I.\]
This right inverse is Schur by Lemma \ref{lem:sz-nagy-foias}.$\hfill\blacksquare$
\end{pf}

\subsection{Minimum Norm Right Inverse}

The extension of Theorem \ref{thm:schur-inverse} from an upper bound
one on right inverses to arbitrary bounds is a simple scaling argument.
Note that in particular the upper bound $\gamma=\gamma_{\opt}(H)$
is valid due to Corollary \ref{cor:min-norm-right-inverse-exists},
and results in a minimum norm right inverse of $H$.

\begin{cor}
Let $\gamma_{\opt}(H)\le\gamma<\infty$. Denote by $\tilde{G}\in S(\mathcal{E}_{*},\mathcal{E})$
the right inverse to $\tilde{H}:=\gamma H$ as given by Theorem \ref{thm:schur-inverse}.
Then $G:=\gamma\tilde{G}$ is a right inverse of $H$ with $\Vert G\Vert_{\infty}\le\gamma$.
\end{cor}
\begin{pf}
Since $\gamma_{\opt}(H)\le\gamma<\infty$, a right inverse $\check{G}\in H^{\infty}(\mathcal{E}_{*},\mathcal{E})$
of $H$ with $\Vert\check{G}\Vert_{\infty}\le\gamma$ exists by Corollary
\ref{cor:min-norm-right-inverse-exists}. Thus\[
\gamma H\gamma^{-1}\check{G}=I,\qquad\Vert\gamma^{-1}\check{G}\Vert_{\infty}\le1,\]
which shows that $\tilde{H}=\gamma H$ has a right inverse in $S(\mathcal{E}_{*},\mathcal{E})$.
Let $\tilde{G}\in S(\mathcal{E}_{*},\mathcal{E})$ denote the right
inverse of $\tilde{H}$ given by Theorem \ref{thm:schur-inverse}.
Then $G=\gamma\tilde{G}$ holds $\Vert G\Vert_{\infty}=\gamma\Vert\tilde{G}\Vert_{\infty}\le\gamma$
as well as\[
HG=\gamma^{-1}\tilde{H}\gamma\tilde{G}=I.\]
$\hfill\blacksquare$
\end{pf}

\section{Conclusions}

In this paper we considered the problem of the construction of a causal
precoder with optimal robustness for a stable and causal LTI system
with multiple inputs and outputs. This problem is equivalent to finding
a solution to the Bezout Identity with minimized peak value, for which
we gave an explicit construction. We derived a novel method for numerical
computation of the lowest peak value achievable in this problem, because
it has to be known prior to the construction of the optimal precoder.
This method is based on computation of a singular value decomposition
of the finite section of a certain infinite block Toeplitz matrix,
which is directly constructed from the Fourier coefficients of the
systems transfer function.

\section*{Appendix}

The complete proof of Theorem \ref{thm:rho-N-converges} follows splitted
in three propositions.

The first proposition shows that $\{\rho_{N}(H)\}_{N}$ is monotonically
decreasing and converges with a limit not lower than $\rho(H)$.

\begin{prop}
\label{pro:rho-k-mon-dec}It holds\[
\rho_{N}(H)\ge\rho_{N+1}(H)\ge\rho(H)\]
for all $N\in\mathbb{N}$.
\end{prop}
\begin{pf}
Let $u\in H^{2}(\mathcal{E}_{*})$. We set $v:=P_{N}u$ and $w:=T_{H^{*}}v$.
A simple computation shows that the Fourier coefficients of $w=P_{+}(H^{*}v)$
are given by\[
\hat{w}_{k}=\left\{ \begin{array}{cl}
\sum_{j=0}^{\infty}\hat{H}_{j}^{*}\hat{v}_{k+j} & ,k\ge0\\
0 & ,k<0\end{array}\right..\]
Since by construction $\hat{v}_{k}=0$ for $k>N$, we see that $\hat{w}_{k}=0$
for $k>N$. Thus\begin{equation}
\Vert T_{H^{*}}P_{N}u\Vert_{2}^{2}=\Vert w\Vert_{2}^{2}=\sum_{k=0}^{\infty}\Vert\hat{w}_{k}\Vert_{2}^{2}=\sum_{k=0}^{N}\Vert\hat{w}_{k}\Vert_{2}^{2}=\Vert P_{N}w\Vert_{2}^{2}=\Vert P_{N}T_{H^{*}}P_{N}u\Vert_{2}^{2}\label{eq:pro:rho-k-mon-dec-1}\end{equation}
holds by Parseval's Relation for every $u\in H^{2}(\mathcal{E}_{*})$.

Because trivially $P_{N}H^{2}(\mathcal{E}_{*})\subset P_{N+1}H^{2}(\mathcal{E}_{*})\subset H^{2}(\mathcal{E}_{*})$,
we obtain with \eqref{eq:pro:rho-k-mon-dec-1}, that\begin{eqnarray*}
\rho_{N}(H) & = & \inf_{u\in P_{N}H^{2}(\mathcal{E}_{*}),\Vert u\Vert_{2}=1}\Vert P_{N}T_{H^{*}}u\Vert_{2}\\
 & = & \inf_{u\in P_{N}H^{2}(\mathcal{E}_{*}),\Vert u\Vert_{2}=1}\Vert T_{H^{*}}u\Vert_{2}\\
 & \ge & \inf_{u\in P_{N+1}H^{2}(\mathcal{E}_{*}),\Vert u\Vert_{2}=1}\Vert T_{H^{*}}u\Vert_{2}\qquad\left(=\rho_{N+1}(H)\right)\\
 & \ge & \inf_{u\in H^{2}(\mathcal{E}_{*}),\Vert u\Vert_{2}=1}\Vert T_{H^{*}}u\Vert_{2}\\
 & = & \rho(H).\end{eqnarray*}

$\hfill\blacksquare$
\end{pf}
We now ensure that the limit of $\{\rho_{N}(H)\}_{N}$ also is not
greater than $\rho(H)$.

\begin{prop}
\label{pro:rho-k-le-rho-plus-eps}For every $\epsilon>0$ there exists
$K\in\mathbb{N}$ such that\[
\rho_{N}(H)\le\rho(H)+\epsilon\]
for all $N>K$.
\end{prop}
\begin{pf}
We assume $H\ne0$ since the case $H=0$ is trivially true. Let $\epsilon>0$
and choose $\check{u}\in H^{2}(\mathcal{E}_{*})$ with $\Vert\check{u}\Vert_{2}=1$
such that\begin{equation}
|\Vert T_{H^{*}}\check{u}\Vert_{2}-\rho(H)|=\left|\Vert T_{H^{*}}\check{u}\Vert_{2}-\inf_{u\in H^{2}(\mathcal{E}_{*}),\Vert u\Vert_{2}=1}\Vert T_{H^{*}}u\Vert_{2}\right|\le\frac{\epsilon}{6}.\label{eq:pro:rho-k-le-rho-plus-eps}\end{equation}
Since $\check{u}\in H^{2}(\mathcal{E}_{*})$, $T_{H^{*}}\check{u}\in H^{2}(\mathcal{E})$
and $\Vert\check{u}\Vert_{2}=1$, Parseval's Relation shows that\[
\lim_{N\to\infty}\Vert P_{N}\check{u}-\check{u}\Vert_{2}=\lim_{N\to\infty}\Vert P_{N}T_{H^{*}}\check{u}-T_{H^{*}}\check{u}\Vert_{2}=0,\qquad\lim_{N\to\infty}\Vert P_{N}\check{u}\Vert_{2}=1.\]
Thus $K\in\mathbb{N}$ exists such that\begin{eqnarray}
\Vert P_{N}\check{u}-\check{u}\Vert_{2} & \le & \frac{\epsilon}{6}\Vert T_{H^{*}}\Vert_{\op}^{-1},\label{eq:pro:rho-k-le-rho-plus-eps-2}\\
\Vert P_{N}T_{H^{*}}\check{u}-T_{H^{*}}\check{u}\Vert_{2} & \le & \frac{\epsilon}{6}\qquad\text{and}\label{eq:pro:rho-k-le-rho-plus-eps-3}\\
\Vert P_{N}\check{u}\Vert_{2} & \ge & \frac{\rho(H)+\frac{\epsilon}{2}}{\rho(H)+\epsilon}\label{eq:pro:rho-k-le-rho-plus-eps-4}\end{eqnarray}
for all $N>K$.

Then for $N>K$ it follows that \begin{eqnarray}
\Vert P_{N}T_{H^{*}}P_{N}\check{u}-T_{H^{*}}\check{u}\Vert_{2} & \le & \Vert P_{N}T_{H^{*}}(P_{N}\check{u}-\check{u})\Vert_{2}+\Vert T_{H^{*}}\check{u}-P_{N}T_{H^{*}}\check{u}\Vert_{2}\nonumber \\
 & \le & \underbrace{\Vert P_{N}T_{H^{*}}\Vert_{\op}}_{\le\Vert T_{H^{*}}\Vert_{\op}}\underbrace{\Vert P_{N}\check{u}-\check{u}\Vert_{2}}_{\le\epsilon/(6\Vert T_{H^{*}}\Vert_{\op})\text{ by }(\ref{eq:pro:rho-k-le-rho-plus-eps-2})}+\underbrace{\Vert T_{H^{*}}\check{u}-P_{N}T_{H^{*}}\check{u}\Vert_{2}}_{\le\epsilon/6\text{ by }(\ref{eq:pro:rho-k-le-rho-plus-eps-3})}\nonumber \\
 & \le & \frac{\epsilon}{3}\label{eq:pro:rho-k-le-rho-plus-eps-5}\end{eqnarray}
and therefore\begin{eqnarray*}
 &  & \left|\Vert P_{N}T_{H^{*}}P_{N}\check{u}\Vert-\inf_{u\in H^{2}(\mathcal{E}_{*}),\Vert u\Vert_{2}=1}\Vert T_{H^{*}}u\Vert_{2}\right|\\
 & \le & \underbrace{|\Vert P_{N}T_{H^{*}}P_{N}\check{u}\Vert_{2}-\Vert T_{H^{*}}\check{u}\Vert_{2}|}_{\le\epsilon/3\text{ by (\ref{eq:pro:rho-k-le-rho-plus-eps-5})}}+\underbrace{\left|\Vert T_{H^{*}}\check{u}\Vert_{2}-\inf_{u\in H^{2}(\mathcal{E}_{*}),\Vert u\Vert_{2}=1}\Vert T_{H^{*}}u\Vert_{2}\right|}_{\le\epsilon/6\text{ by }(\ref{eq:pro:rho-k-le-rho-plus-eps})}\\
 & \le & \frac{\epsilon}{2}.\end{eqnarray*}
We see that\begin{equation}
\Vert P_{N}T_{H^{*}}P_{N}\check{u}\Vert_{2}\le\inf_{u\in H^{2}(\mathcal{E}_{*}),\Vert u\Vert_{2}=1}\Vert T_{H^{*}}u\Vert_{2}+\frac{\epsilon}{2}=\rho(H)+\frac{\epsilon}{2}.\label{eq:eq:pro:rho-k-le-rho-plus-eps-6}\end{equation}

Since $\Vert P_{N}\check{u}\Vert_{2}>0$ for $N>K$ by \eqref{eq:pro:rho-k-le-rho-plus-eps-4},
the sequence $\{\check{u}_{N}\}_{N>K}$ given by\[
\check{u}_{N}:=\frac{P_{N}\check{u}}{\Vert P_{N}\check{u}\Vert_{2}}\in P_{N}H^{2}(\mathcal{E}_{*})\]
is well-defined. We obtain the intended result\begin{eqnarray*}
\rho_{N}(H) & = & \inf_{u\in P_{N}H^{2}(\mathcal{E}_{*}),\Vert u\Vert_{2}=1}\Vert P_{N}T_{H^{*}}u\Vert_{2}\\
 & \le & \Vert P_{N}T_{H^{*}}\check{u}_{N}\Vert_{2}\\
 & = & \frac{\Vert P_{N}T_{H^{*}}P_{N}\check{u}\Vert_{2}}{\Vert P_{N}\check{u}\Vert_{2}}\\
 & \stackrel{\text{(by (\ref{eq:eq:pro:rho-k-le-rho-plus-eps-6}))}}{\le} & \frac{\rho(H)+\frac{\epsilon}{2}}{\Vert P_{N}\check{u}\Vert_{2}}\\
 & \stackrel{\text{(by (\ref{eq:pro:rho-k-le-rho-plus-eps-4}))}}{\le} & \rho(H)+\epsilon\end{eqnarray*}
for all $N>K$.$\hfill\blacksquare$
\end{pf}
We know now by the Propositions \ref{pro:rho-k-mon-dec} and \ref{pro:rho-k-le-rho-plus-eps}
that the sequence $\rho_{N}$ converges to $\rho$ for $N\to\infty$.
However it is still unclear, how $\rho_{N}$ can be computed explicitly.
The next proposition gives a simple formula for the numerical computation
of $\rho_{N}$.

\begin{prop}
\label{pro:rho-N-is-sigma-min}Let $H\in H^{\infty}(\mathbb{C}^{m\times n})$
with $m\le n$ and set \[
\Gamma_{H,N}:=\left[\begin{array}{cccc}
\hat{H}_{0}^{*} & \hat{H}_{1}^{*} & \dots & \hat{H}_{N}^{*}\\
0 & \hat{H}_{0}^{*} & \dots & \hat{H}_{N-1}^{*}\\
\vdots & \ddots & \ddots & \vdots\\
0 & \dots & 0 & \hat{H}_{0}^{*}\end{array}\right]\in\mathbb{C}^{n(N+1)\times m(N+1)}.\]
 Then $\rho_{N}(H)=\sigma_{\min}(\Gamma_{H,N})$.
\end{prop}
\begin{pf}
Let $USV^{*}=\Gamma_{H,N}$ denote a singular value decomposition
of $\Gamma_{H,N}$ with singular values\[
\sigma_{1}\ge\dots\ge\sigma_{m(N+1)}\ge0.\]
Then $U\in\mathbb{C}^{n(N+1)\times n(N+1)}$ and $V\in\mathbb{C}^{m(N+1)\times m(N+1)}$
are unitary matrices and $S\in\mathbb{C}^{n(N+1)\times m(N+1)}$ is
of the form\[
S=\left[\begin{array}{ccc}
\sigma_{1}\\
 & \ddots\\
 &  & \sigma_{m(N+1)}\\
\\\\\end{array}\right].\]

Let $u\in P_{N}H^{2}(\mathbb{C}^{n})$ and set $v:=P_{N}T_{H^{*}}u$.
We saw already in the proof of Proposition \ref{pro:rho-k-mon-dec},
that the non-zero Fourier coefficients of $v$ are uniquely determined
by the relation\[
\left[\begin{array}{c}
\hat{v}_{0}\\
\hat{v}_{1}\\
\vdots\\
\hat{v}_{N}\end{array}\right]=\left[\begin{array}{cccc}
\hat{H}_{0}^{*} & \hat{H}_{1}^{*} & \dots & \hat{H}_{N}^{*}\\
0 & \hat{H}_{0}^{*} & \dots & \hat{H}_{N-1}^{*}\\
\vdots & \ddots & \ddots & \vdots\\
0 & \dots & 0 & \hat{H}_{0}^{*}\end{array}\right]\left[\begin{array}{c}
\hat{u}_{0}\\
\hat{u}_{1}\\
\vdots\\
\hat{u}_{N}\end{array}\right]=\Gamma_{H,N}\left[\begin{array}{c}
\hat{u}_{0}\\
\hat{u}_{1}\\
\vdots\\
\hat{u}_{N}\end{array}\right].\]
Thus by Parseval's Relation \begin{eqnarray*}
\rho_{N}(H) & = & \inf_{u\in P_{N}H^{2}(\mathbb{C}^{m}),\Vert u\Vert_{2}=1}\Vert P_{N}T_{H^{*}}u\Vert_{2}\\
 & = & \inf_{u\in\mathbb{C}^{m(N+1)},\Vert u\Vert_{2}=1}\Vert\Gamma_{H,N}u\Vert_{2}\\
 & = & \inf_{u\in\mathbb{C}^{m(N+1)},\Vert u\Vert_{2}=1}\Vert Su\Vert_{2}\\
 & = & \sigma_{m(N+1)}\\
 & = & \sigma_{\min}(\Gamma_{H,N}).\end{eqnarray*}
$\hfill\blacksquare$
\end{pf}
\bibliographystyle{elsart-num}
\bibliography{/home/sander/text/bibtex/database}

\end{document}